\documentclass[aps,prd,preprint,nofootinbib]{revtex4}
\usepackage{amsfonts}
\usepackage{mathrsfs}
\usepackage{graphicx}
\usepackage{amsmath}
\usepackage{amssymb}
\usepackage{subfigure}
\usepackage{epsfig}
\usepackage{graphicx}
\usepackage{slashed}
\usepackage{color}
\usepackage{float}
\newcommand{\bqa}{\begin{eqnarray}}
\newcommand{\eqa}{\end{eqnarray}}
\newcommand{\beq}{\begin{equation}}
\newcommand{\eeq}{\end{equation}}
\allowdisplaybreaks[1]
\graphicspath{{fig/}{dia/}} \DeclareGraphicsExtensions{.eps}
\begin{document}

\title{The newly observed state $D_{s0}(2590)^{+}$ \\[0.7cm]}

\author{\vspace{1cm}Guo-Li Wang$^{1,2}$\footnote[1]{wgl@hbu.edu.cn, corresponding author}, Wei Li$^{1,2}$, Tai-Fu Feng$^{1,2}$\footnote[1]{fengtf@hbu.edu.cn, corresponding author},Ying-Long Wang$^{1,3}$, Yu-Bin Liu$^{4}$}

\affiliation{$^1$ Department of Physics, Hebei University, Baoding 071002, China\\
$^2$ Key Laboratory of High-precision Computation and Application of Quantum Field Theory of Hebei Province, Baoding, China\\
$^3$ Department of Primary Education, Baoding Preschool Teachers College, Baoding 072750, China\\
$^4$ School of Physics, Nankai University, Tianjin 300071, China\vspace{0.6cm}}

\begin{abstract}
\vspace{0.5cm}
We choose the Reduction Formula, PCAC and Low Energy Theory to reduce the $S$ matrix of a OZI allowed two-body strong decay involving a light pseudoscalar, the covariant transition amplitude formula with relativistic wave functions as input is derived. After confirm this method by the decay $D^*(2010)\to D\pi$, we study the newly observed $D_{s0}(2590)^{+}$ with supposing it to be the state $D_s(2^1S_0)^+$, we find its decay width $\Gamma$ is highly sensitive to the $D_{s0}(2590)^{+}$ mass, which result in the meaningless comparison of widths by different models with various input masses. Instead of width, we studied the overlap integral over the wave functions of initial and final states, here we parameterized it as $X$ which is model-independent, and the ratio $\Gamma/{|{\vec P_f}|^3}$, both are almost mass independent, to give us useful information. The results show that, all the existing theoretical predictions  $X_{D_s(2S) \to D^*K}=0.25\sim 0.41$ and $\Gamma/{|{\vec P_f}|^3}=0.81\sim1.77$ MeV$^{-2}$ are smaller than experimental data $0.585^{+0.015}_{-0.035}$ and $4.54^{+0.25}_{-0.52}$ MeV$^{-2}$. Further compared with $X^{ex}_{D^*(2010) \to D\pi}=0.540\pm0.009$, the current data $X^{ex}_{D_s(2S) \to D^*K}=0.585^{+0.015}_{-0.035}$ is too big to be an reasonable value, so it is early to say $D_{s0}(2590)^{+}$ is the conventional $D_s(2^1S_0)^+$ meson.

\end{abstract}
 \maketitle
\newpage

\section{Introduction}\label{Sec-1}

In recent years, great progress in the mass spectra of charmed and charmed-strange mesons has been made in experiments, many excited states are observed \cite{Chen:2016spr}, for example, $D(2550)$ was observed in the $D^*\pi$ mass distribution by the BaBar Collaboration in 2010 \cite{delAmoSanchez:2010vq}, though there are some disagreements \cite{liux2,zhongxh}, it is a good candidate for $D(2^1S_0)$ \cite{GM,zhangal2}, the first radial excited state of the $0^-$ pseudoscalar $D(1^1S_0)$. Three years later, the LHCb Collaboration reported the $D_J(2580)$ in $D^*\pi$ invariant mass spectrum \cite{Aaij:2013sza}, since they have similar properties, $D(2550)$ and $D_J(2580)$ may be the same particle. For the vector excited $1^-$ state $D^*(2^3S_1)$, there are three candidates, $D^*(2600)$, $D_J^*(2650)$ and $D_1^*(2680)^0$, observed by BaBar \cite{delAmoSanchez:2010vq} and LHCb Collaborations \cite{Aaij:2013sza,Aaij:2016fma}, respectively.

For the charm-strange meson, in the year 2004, $D^*_{s}(2632)$, as the candidate of the first radial excited $1^-$ state, was reported by the SELEX Collaboration in invariant mass spectra of $D^+_s\eta$ and $D^0K^+$ \cite{Evdokimov:2004iy}.  Theoretically, by using the Reduction Formula, the Partial Conservation of the Axial Current (PCAC), the Low Energy Theory, and solved the instantaneous Bethe-Salpeter equation, we studied the mass and Okubo-Zweig-Iizuka (OZI) allowed two-body strong decays of $D^*_{s}(2^3S_1)$. In contrast to data, we obtained a higher mass and a broader width, we drew a conclusion that it is too early to conclude that $D^*_{s}(2632)$ is the first radial excitation of the $D_s^*(2112)$ \cite{2632}. There are also many theoretical studies disfavor this assumption \cite{zhusl,chenyq,barnes,daiyb,close}. Up to now, this narrow state did not confirmed by other experiments. In the year 2006, a broad structure named as $D^*_{s1}(2700)$ was observed by the BaBar Collaboration in the $DK$ invariant mass spectrum \cite{Aubert:2006mh}, and it was confirmed by Belle \cite{Brodzicka:2007aa}, BaBar \cite{Aubert:2009ah,Lees:2014abp} and LHCb \cite{Aaij:2012pc} experiments. This $1^-$ state $D^*_{s1}(2700)$ is a good candidate of the radial excited state $D^*_{s}(2^3S_1)$ \cite{close2,GM}.

Recently, using $pp$ collision data collected with the LHCb detector at a centre-of-mass energy of 13 TeV, the $B^0\to D^+D^-K^+\pi^-$ decay is studied, a new state named $D_{s0}(2590)^+$ is observed \cite{Aaij:2020voz} in the $D^+K^+\pi^-$ invariant mass spectrum, whose mass and decay width are detected as $m=2591\pm6\pm7$ MeV and $\Gamma=89\pm16\pm12$ MeV. Since it decays into the $D^+K^+\pi^-$ final state, its spin-parity are measured with an amplitude analysis, and its $J^P=0^-$ is confirmed, since the only missing low excited charm-strange meson is the pseudoscalar $2^1S_0$ state, so $D_{s0}(2590)^+$ is believed to be a strong candidate of the $D_{s}(2^1S_0)^+$ state.


\begin{figure}[H]
\begin{picture}(470,120)(70,540)
\put(0,0){\includegraphics{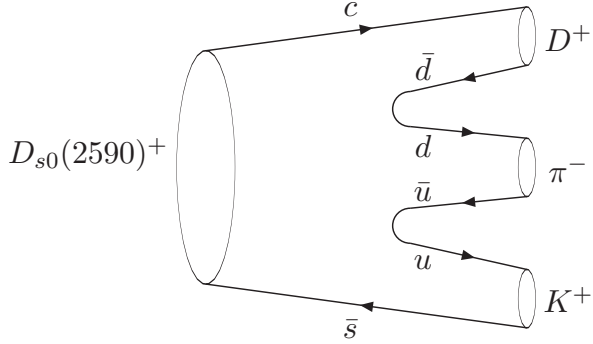}}
\end{picture}
\caption{The detected channel $D_{s0}(2590)^+\to D^+K^+\pi^-$ by LHCb.
}\label{figure1}
\end{figure}

In the discovery experiment,
the detected channel is $D_{s0}(2590)^+\to D^+K^+\pi^-$,  see Figure 1, it's an OZI allowed three-body strong decay (not a cascade decay of a two-body strong process), but not the dominant decay of $D_{s0}(2590)^+$ as the state $D_{s}(2S)^+$ because there are OZI allowed two-body strong decays, for example, the decay channel shown in Figure 2. Compared with two-body decay, this three-body process suffer from both the phase space and QCD suppressions, so instead of the three-body channels, such OZI allowed two-body strong decays play an important role in determining the property of this particle, for example, it can be used to roughly estimate the full width.

\begin{figure}[H]
\begin{picture}(470,120)(70,530)
\put(0,0){\includegraphics{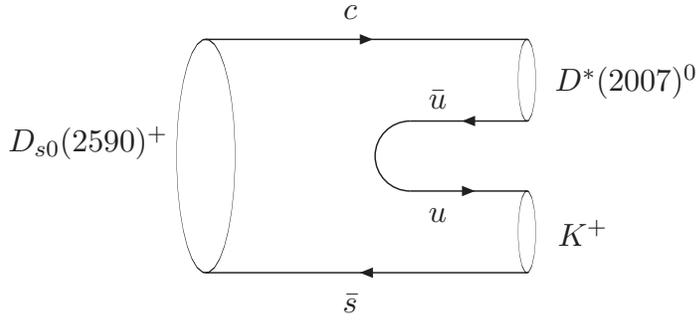}}
\end{picture}
\caption{Dominant decay channel $D_{s0}(2590)^+\to D^*(2007)^0K^+$ ($D_{s0}(2590)^+\to D^*(2010)^+K^0$ when $u\bar u$ is changed to $d \bar d$).
}\label{figure2}
\end{figure}

As a $J^P=0^-$ state, its possible strong decay channels are $0^-\to 1^-0^-$, $ 1^-1^-$, $0^-0^+$, $1^-0^+$ and $0^-1^+$, etc, but limited by the mass threshold, the $DK^*$, $D^*_s\eta$ and other channels are forbidden, only two channels survive, they are $D_{s0}(2590)^+\to D^*(2007)^0K^+$ and $D_{s0}(2590)^+\to D^*(2010)^+K^0$.

There are already some theoretical predictions of the two-body strong decays of $D_s(2^1S_0)$ using different models, for example, Ref. \cite{GM} used the relativized quark model and $^3P_0$ quark pair creation model; Ref. \cite{liux} chose the Godfrey-Isgur model and $^3P_0$ model; Refs. \cite{close,zhangal} chose the harmonic oscillator wave functions and $^3P_0$ model; Ref.\cite{colangelo} adopted an effective Lagrangian approach based on the heavy quark and chiral symmetry; our previous study \cite{wangzh} chose the Reduction Formula and the PCAC to simplify the transition matrix element, then adopted two methods to make further calculations, first one is the Low Energy Theory, another one is the Impulse Approximation \cite{roberts}, both of them used the relativistic wave functions by solving the instantaneous Bethe-Salpeter equation.

In this paper, we will revisit the topic of $D_s(2^1S_0)$, and study the possibility of $D_{s0}(2590)^+$ as the $D_s(2^1S_0)$. The reason is that, first, the detected mass of $D_{s0}(2590)^+$ is smaller than all the theoretical predictions about $D_{s}(2^1S_0)$, at least several tens of MeV smaller; second, all the calculations of decay width based on a much higher $D_s(2^1S_0)$ mass. At first sight, it seems some theoretical predictions of width consist with data, but we point out that it is not true, with different masses as input, the comparison of decay widths is meaningless because the OZI allowed strong decays happen closing to the mass threshold of $D_{s}(2^1S_0)$, which make the width highly sensitive to the input mass. So to compare the width with experimental data we need to do the calculation using the same mass with data.

As an alternative, the ratio $\Gamma/{|{\vec P_f}|^3}$ \cite{colangelo} ( $\Gamma$ and ${\vec P_f}$ are the width and recoil momentum, respectively) can cancel partly the influence of different input masses.
We further study the overlap integral over the initial and final state wave functions, which is parameterized as a model independent quantity $X$.
The quantity $X$ remove the effect of mass to a great extent, and make all the theoretical calculations and the experimental data are comparable no matter what the $D_{s}(2^1S_0)$ mass is. In another word, we do not need to recalculate the strong decays with same mass as input, but only to compute the quantity $X$ using the existing width result. Since $X$ is overlap integral, so its value can only be within a reasonable range, then from its value we can draw a conclusion.

In our method, we will choose the Reduction Formula, PCAC as well as the Low Energy Theory, but with an improved more covariant hadronic transition amplitude formula, where the input relativistic wave functions are obtained by solving the Salpeter equations. The Bethe-Salpeter (BS) equation \cite{BS}, based on the quantum field theory, is a relativistic dynamic equation describing bound state. Salpeter equation \cite{Sal} is its instantaneous version, and is suitable for heavy mesons. To confirm our method, we first study the strong decays of $D^{*}(2010)^+$, which are already well measured in experiment, then we apply this method to the study of $D_{s0}(2590)^+$.

This paper is organized as followings,
in Sec. II, we summarize the theoretical predictions of $D_s(2^1S_0)$ mass and the mass splittings in experimental data, and give our comment; in Sec. III, the formula of relativistic transition amplitude and the relativistic wave functions are presented. In Sec. IV, we make non-relativistic limit of our method, then introduce the model independent quantity $X$; the numerical results and discussions are shown in Sec. V. Finally, we show the detail of deriving the covariant transition amplitude with instantaneous wave functions as input in Appendix.

\section{The Mass of $D_s(2^1S_0)$}

The mass of $D_{s}(2^1S_0)^+$ has been studied theoretically by many models, we list some of them in Talbe I. We note that, the detected mass $m=2591\pm6\pm7$ MeV of $D_{s0}(2590)^+$ as the $D_{s}(2^1S_0)^+$ candidate is lower at least several tens of MeV than all the theoretical predictions. To compare the results, the mass splitting is more convenient than the mass itself, so the corresponding hyperfine splitting $\Delta M =M_{D_{s}(2 ^1S_{0})}-M_{D_{s}(1 ^1S_{0})}$ is also shown in Table I, where we can see that all the theoretical predictions of $\Delta M$, including the smallest $\Delta M =670$ MeV, are larger than experimental data  $\Delta M =623\pm13$ MeV. Similar thing happens to the case of $D^{*}_s(2632)$, whose mass is detected as $2632.5\pm1.7$ MeV which is smaller than all the theoretical predictions of $D^{*}_{s}(2^3S_1)^+$, currently the experimental average mass of $D^{*}_{s}(2^3S_1)^+$ is $2708^{+4.0}_{-3.4}$ \cite{pdg}, which is consistent with most of the theoretical predictions.

\begin{table}
\begin{center}
\caption{Masses of $D_{s}(1 ^1S_{0})$, $D_{s}(2 ^1S_{0})$ and their mass splitting  in unit of MeV.}
\label{tab1}
\begin{tabular}{|c|c|c|c|c|c|c|c|c|c|c|}\hline\hline
& \cite{GI,GM} &\cite{Ebert}& \cite{lidemin}&\cite{Eichten}&\cite{Lahde}&\cite{Zeng}&\cite{liux}&Ex~\cite{pdg,Aaij:2020voz}\\ \hline						
~$M(1 ^1S_{0})$   ~& 1979 &~ 1969 ~& ~1969~ &~1965~ &~1975~&~1940~&~1967~&~1968.30$\pm$0.11~ \\\hline
~$M(2 ^1S_{0})$  ~ &~ 2673~ &~ 2688~ &~ 2640~ &~2700~ &2659&2610&2646&2591$\pm$13\\\hline
~$\Delta M(2 ^1S_{0}-1 ^1S_{0})$  ~ &~ 694~ &~ 719~ &~ 671~ &~735~ &~684~&670&679&623$\pm$13\\\hline
\hline
\end{tabular}
\end{center}
\end{table}

There are other arguments which can help us to test the mass of $D_{s}(2^1S_0)^+$. In Table II, we list some mass splittings based on the experimental data, where the large uncertainties come from the following newly observed hadrons, $D(2S)$, $D^*(2S)$ and $D^*_s(2S)$, their masses are  $M_{D(2550)^0}=2564\pm 20$ MeV, $M_{D^*_J(2600)}=2623\pm 12$ MeV, and $M_{D^*_{s1}(2700)^+}=2708^{+4.0}_{-3.4}$ MeV \cite{pdg}. These three states are also not well measured, but each of them has several experimental detections, so the mass information of $D_s(2S)$ can be roughly extracted from these states and other well established mesons.

In Table \ref{splitting}, the values of mass splitting in the first two columns are relatively well measured experimentally, and from the first row to the third row, these values approximately satisfy a decreasing trend. We assume that the fourth column meets the rule of the first column, and the third column has the same rule as the second column, then we approximately have $\Delta M(c\bar u)>\Delta M(c\bar s)>\Delta M(c\bar c)$.
But the current values $\Delta M(2^1S_0-1^1S_0)\equiv$$M(2^1S_0)-M(1^1S_0)=623\pm13$ MeV and $ M(2^3S_1)-M(2^1S_0)=117^{+17}_{-16}$ MeV, see the second row for the $c\bar s$ system in Table \ref{splitting}, where $D_{s0}(2590)^+$ is treated as the $D_s(2^1S_0)$, conflict with this roughly decreasing rule. For the last two columns, since compared with the second row, the values in the first and third rows are relatively well measured, so we can use them and the decreasing rule to estimate the values in the second row. According to the decreasing rule, the third column shows that the mass of $D_s(2^1S_0)$ should be between $2620$ and $2687$ MeV, and the fourth column indicates its mass should be between $2614$ and $2665$ MeV, combine them, the mass of $D_s(2^1S_0)$ should be located at $2620 \to 2665$ MeV, the current mass $2591\pm13$ MeV is lower than this expectation.

\begin{table}
\begin{center}
\caption{Mass splittings (MeV) based on the data in PDG \cite{pdg} and Ref.\cite{Aaij:2020voz}.}
\label{splitting}
\begin{tabular}{|c|c|c|c|c|}\hline\hline
 &~~$\Delta M(1^3S_1-1^1S_0)$~~&~~$\Delta M(2^3S_1-1^3S_1)$~~&~~$ \Delta M(2^1S_0-1^1S_0)$~~&~~$\Delta M(2^3S_1-2^1S_0)$~~ \\\hline
~~$c\bar u$~~ &$142.0\pm0.1$&$616\pm 12$&$699\pm 20$&$59\pm 32$ \\\hline
~~$c \bar s$~~ &$143.9\pm 0.5$&$596.1^{+4.4}_{-3.8}$&`~$623\pm 13$~'&`~$117^{+17}_{-16} $~'\\\hline
~~$c\bar c$~~ &$113.0\pm 0.5$&$589.20\pm 0.07$&$653.6\pm 1.6$&$48.6\pm 1.2$ \\\hline
 \hline
\end{tabular}
\end{center}
\end{table}

\section{Transition $S$ matrix and decay width}\label{Sec-2}

As the radial excited $0^-$ state, $D_{s0}(2590)^{+}$ has two OZI allowed strong decay channels, $D_{s0}(2590)^{+} \to D^{*}(2007)^0+K^{+}$ and $D_{s0}(2590)^{+} \to D^{*}(2010)^{+}+K^0$, the corresponding Feynman diagrams are shown in Figure 2. Considering such a diagram, the $^3P_0$ model \cite{3p0,3p01} is widely used to calculate such kind of decays, and the transition amplitude is written as overlapping integral over the non-relativistic wave functions of the corresponding initial and final mesons. Since $K$ is a light meson, its non-relativistic wave function may bring large uncertainty, so we abandon the $^3P_0$ model.

To give a rigorous calculation, we adopt the Reduction Formula to avoid using non-relativistic $K$ meson wave function, then the transition $S$-matrix for the decay $D_{s0}(2590)^{+} \to D^{*}K$ can be written as,
\begin{equation}
\label{S matrix}
\begin{split}
\langle {D^*}(P_f)K(P_{f2})|D_{s0}(P)^+\rangle=\int d^4x e^{iP_{f2}\cdot x}
(M^2_{K}-P^2_{f2})\langle D^*(P_{f})|\phi_K(x)|D_{s0}(P)^+\rangle,
\end{split}
\end{equation}
where $\phi_K$ is the field of $K$ meson, which can be related to the axial current because of the PCAC,
$\phi_K(x)=\frac{1}{M^2_K f_K}\partial^{\mu}(\bar{q}\gamma_{\mu}\gamma_5 s)$, where $q=u,d$ for $K^+,K^0$, respectively, and $f_K$ is the decay constant of $K$ meson. Using the integration by parts, we obtain the following relation,
$$
\int d^4x e^{iP_{f2}\cdot x}
\langle D^*(P_{f})|\partial^{\mu}(\bar{q}\gamma_{\mu}\gamma_5 s)|D_{s0}(P)^+\rangle=-iP^{\mu}_{f2}\int d^4x e^{iP_{f2}\cdot x}
\langle D^*(P_{f})|\bar{q}\gamma_{\mu}\gamma_5 s|D_{s0}(P)^+\rangle.
$$
Since the mass of $D_{s0}(2590)^{+} $ is just above the threshold of $D^{*}K$, the Low Energy Theory indicate that $P^2_{f2}\to 0$, finally after the integral over $x$, the $S$-matrix becomes \cite{2632}
$$\langle {D^*}(P_f)^0K(P_{f2})|D_{s0}(P)^+\rangle=
(2\pi)^4\delta^4(P-P_f-P_{f2})\mathcal{M}$$
\begin{equation}
\label{amplitude}
\begin{split}
=-(2\pi)^4\delta^4(P-P_f-P_{f2})\frac{iP^{\mu}_{f2}}{f_K}\langle {D^*}(P_f)^0|\bar{q}\gamma_{\mu}\gamma_5s|D_{s0}(P)^+\rangle,
\end{split}
\end{equation}
where $\mathcal{M}$ is the transition amplitude.
In this case, the Feynman diagram in Figure 2 can be reduced to the one drew in Figure 3.
\begin{figure}[H]
\begin{picture}(470,150)(70,510)
\put(0,0){\includegraphics{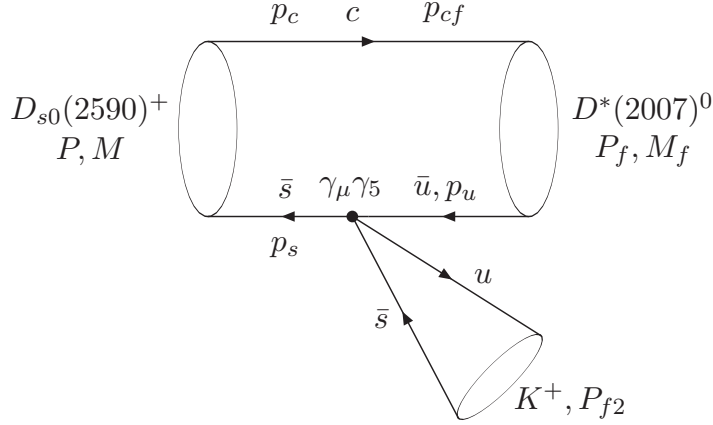}}
\end{picture}
\caption{Feynman diagram for decay $D_{s0}(2590)^+\to D^*(2007)^0K^+$ after reduction.
}\label{figure3}
\end{figure}

In the previous study \cite{2632}, the transition amplitude is written as overlapping integral over the positive wave functions ${\varphi}^{++}_{_{P}}({q}_{_{P\bot}})$ and $\bar{\varphi}^{++}_{_{P_f}}({q_{f}}_{_{P\bot}})$ of initial and final mesons \cite{Chang:2006tc},
\begin{equation}\label{oldamp}
\langle {D^*}(P_f)^0|\bar{q}\gamma_{\mu}\gamma_5s|D_{s0}(P)^+\rangle=\int \frac{\mathrm{d}^{3} q_{_{P\bot}}}{(2 \pi)^{3}} \operatorname{Tr}\left[\bar{\varphi}^{++}_{_{P_f}}({q_{f}}_{_{P\bot}})
{\frac{\not\!P}{M}}
{\varphi}^{++}_{_{P}}({q}_{_{P\bot}})
 \gamma_{\mu}\gamma_5\right],
\end{equation}
where the relative momentum $q_{_{P\perp}}= q-q_{_P}\frac{P}{M}$ with $q_{_P}=\frac{P\cdot q}{M}$, and
the relation ${q_{f}}_{_{P\bot}}={q}_{_{P\bot}}-\alpha'_c {P_f}_{_{P\bot}}~(\alpha'_c=\frac{m_c}{m_c+m_u})$ in final state wave function is widely used in literature, but since the wave function of final state is solved in its own center of mass system, so instead we use a more covariant expression for the transition amplitude
\begin{equation}\label{amplitude1}
\int \frac{\mathrm{d}^{3} q_{_{P\bot}}}{(2 \pi)^{3}} \operatorname{Tr}\left[{\Lambda^+(-p_{u_{_{P\perp}}})}
{\frac{\not\!P_f}{M_f}}\frac{
M_f-\widetilde\omega_c-\widetilde\omega_{u}}
{E_f-\omega_c-\omega_{u}}
\right.\bar{\varphi}^{++}_{_{P_f}}({q_{f}}_{_{P_f\bot}})\left.
{\frac{\not\!P_f}{M_f}}
{\varphi}^{++}_{_{P}}({q}_{_{P\bot}})
 \gamma_{\mu}\gamma_5\right],
\end{equation}
where we have the relation
$$
{q_{f}}_{_{P_f\bot}}={q_{f}}_{_{P\bot}}-\frac{{q_{f}}_{_{P\bot}}\cdot {P_f}_{_{P\bot}}}{M^2_f}P_f+(\frac{m_u}{m_c+m_u}\frac{P_f\cdot P}{M}-\omega_u)\left(\frac{P}{M}-\frac{P_f\cdot P}{MM^2_f}P_f\right).
$$
The detail of how to give the Eq.(\ref{amplitude1}) and explanations of various quantities are shown in Appendix.

The general relativistic wave function for a $0^-$ state, for example $D_{s0}(2590)^{+}$, in the condition of instantaneous approximation ($P\cdot q=0$) can be written as \cite{Kim:2003ny},
\begin{equation}
\begin{aligned}\label{pseudowave}
\varphi_{_P}({q}_{_{P\bot}})=\displaystyle
\left(f^{}_1 M+f^{}_2\not\!P+f_3\not\!{q}_{_{P\bot}}
+f_4\frac{\not\!{q}_{_{P\bot}}\not\!P}{M}\right)\gamma^{5},
\end{aligned}
\end{equation}
where $f_i~(i=1,2,3,4)$ is the radial part of the wave function,  and its numerical value is achieved by solving the full Salpeter equation \cite{Kim:2003ny} for a $0^-$ state.

The relativistic wave function of a
vector $1^{-}$ state can be written as \cite{Wang:2005qx}:
$$\varphi_{P_{_f}}({q_{f}}_{_{P_f\bot}})=
{q_{f}}_{_{P_f\bot}}\cdot{\epsilon}
\left[g_1+\frac{\not\!P_f}{M_f}g_2+
\frac{{\not\!{q_{f}}_{_{P_f\bot}}}}{M_f}g_3+\frac{{\not\!P_f}
{\not\!{q_{f}}_{_{P_f\bot}}}}{M_f^2} g_4\right]+
M_f{\not\!\epsilon}g_5$$
\begin{equation}\label{vecwave}+
{\not\!\epsilon}{\not\!P_f}g_6+
({\not\!{q_{f}}_{_{P_f\bot}}}{\not\!\epsilon}-
{q_{f}}_{_{P_f\bot}}\cdot{\epsilon})
g_7+\frac{1}{M_f}({\not\!P_f}{\not\!\epsilon}
{\not\!{q_{f}}_{_{P_f\bot}}}-{\not\!P_f}{q_{f}}_{_{P_f\bot}}\cdot{\epsilon})
g_8,
\end{equation}
where ${\epsilon}$ is the polarization
vector of the meson. The numerical values of the 8 radial wave functions $g_i~(i=1,2,..8)$ are obtained by solving the corresponding Salpeter equation for a $1^-$ state \cite{Wang:2005qx}.
Since the corresponding positive wave functions are obtained straightforward, we will not show them here,  interested readers can find them in Refs. \cite{Kim:2003ny,Geng}.

The two-body decay width is
\begin{equation}\label{gamma}
\Gamma=\frac{|\vec{P_f}|}{8\pi M^2}\frac{1}{2J+1}\sum |\mathcal{M}|^2,
\end{equation}
where if the final state is $\pi^0$ instead of $\pi^+$, there is an extra parameter $1/2$;
$J$ is the total spin of the initial meson; $\vec{P_f}$ is the three-dimension recoil momentum of the final meson
$|\vec{P_f}|=\sqrt{[M^2-(M_f-M_{f2})^2][M^2-(M_f+M_{f2})^2]}/{(2M)}.$
Eq.(\ref{gamma}) shows that $\Gamma \varpropto |\vec{P_f}|$, but since $\mathcal{M} \varpropto P\cdot \epsilon$, and $\sum |P\cdot \epsilon|^2=\frac{M^2 \vec{P_f}^2}{M^2_f}$, so actually we have $\Gamma \varpropto |\vec{P_f}^3|$, means that $\Gamma$ is very sensitive to the value of recoil momentum $\vec{P_f}$. Value $|\vec{P_f}|$ is determined by the initial and final state masses, since two final states are both well established and their masses are well measured, only initial state mass is not well measured and with large errors. We also note that a large mass $M$ will result in a large value $|\vec{P_f}|$, so in another word, the $\Gamma$ is very sensitive to the value of initial meson mass, then the ratio $\Gamma / |\vec{P_f}^3|$ can cancel partly the influence of initial state mass.

\section{A model independent quantity $X$}
We have shown that the decay width is very sensitive to the value of initial state mass, since the OZI allowed decay happens closing to the mass threshold, this strengthen the sensitivity of dependence on mass value. There are some theoretical predictions of decay width by different models but with various masses, which make these theoretical results are incomparable, so removing the mass dependence is crucial.

To realize this purpose, we like to show the non-relativistic limit of our calculation. In the non-relativistic limit, the wave function Eq.(\ref{pseudowave}) of a pseudoscalar becomes
\begin{equation}
\begin{aligned}
\varphi_{_P}^{0^-}({q}_{_{P\bot}})=\displaystyle
\left(M+\not\!P\right)\gamma^{5}~f_1({q}_{_{P\bot}}),
\end{aligned}
\end{equation}
and the wave function Eq.(\ref{vecwave}) for a vector becomes
\begin{equation}\varphi^{1^{-}}_{P_{_f}}({q_{f}}_{_{P_f\bot}})=
\left( M_f+{\not\!P_f} \right)
{\not\!\epsilon} ~ g_5 ({q_{f}}_{_{P_f\bot}}),
\end{equation}
in this case, the normalization conditions are
\begin{equation}\label{nor}
4M\int f_1^2~\frac{d^3{q}_{_{P\bot}}}{(2\pi)^3}\equiv \int {f'}_1^2~\frac{d^3{q}_{_{P\bot}}}{(2\pi)^3}=1,
\end{equation}
\begin{equation}\label{nor1}
4M_f\int g_5^2~\frac{d^3{q_{f}}_{_{P_f\bot}}}{(2\pi)^3}\equiv\int {g'}_5^2~\frac{d^3{q_{f}}_{_{P_f\bot}}}{(2\pi)^3}=1,
\end{equation}
where we have redefine two mass independent wave functions ${f'}_1({q}_{_{P\bot}})$ and ${g'}_5({q_{f}}_{_{P_f\bot}})$.
In this non-relativistic limit, we choose the old previous amplitude formula Eq.(\ref{oldamp}) to do the calculation, then the decay width for channel $i$ is obtained
\begin{equation}\label{nonamp}
\Gamma_i=\frac{{\vec{P_f}}^3(M+M_f)^2}{8\pi f^2_K M M_f}
\left[\int {f'}_1({q}_{_{P\bot}})~{g'}_5({q_{f}}_{_{P\bot}})~\frac{d^3 {q}_{_{P\bot}}}{(2\pi)^3} \right]^2
\equiv \frac{{\vec{P_f}}^3(M+M_f)^2}{8\pi f^2_K M M_f}~X_i^2,
\end{equation}
where we define a quantity
\begin{equation}\label{X}
X_i=\int {f'}_1({q}_{_{P\bot}})~{g'}_5({q_{f}}_{_{P\bot}}) ~\frac{d^3 {q}_{_{P\bot}}}{(2\pi)^3},
\end{equation}
which is the overlapping integral over the initial and final meson wave functions, since the wave functions themselves are mass independent shown in normalization conditions Eq.(\ref{nor}) and Eq.(\ref{nor1}), so the quantity $X_i$ is almost free from mass. But we should point out that, $X_i$ is still slightly dependent on the meson masses, because in the overlapping integral Eq.(\ref{X}) where the internal momentum ${q_{f}}_{_{P\bot}}={q}_{_{P\bot}}-\alpha'_c {P_f}_{_{P\bot}}$ (that is ${\vec q_f}=\vec{q}-\alpha'_c {\vec P_f}$) is used, and the recoil momentum  ${\vec P_f}$ is related to initial and final masses.

Regarding the decay width formula in non-relativistic limit, we notice that,
using heavy quark effective theory, Wang \cite{wangzg} also obtained a similar relation, that is, the dependence of the decay width on masses $M$, $M_f$ and momentum $\vec{P_f}$ in Eqs.(15-16) in Ref. \cite{wangzg} is exactly the same as ours in Eq.(\ref{nonamp}), also in later formulas Eqs.(\ref{18}-\ref{19}) in this paper, this could be a good cross-checked confirmation of the correctness of the decay width formula in two different methods.

From the definition equation Eq.(\ref{X}) and the normalization condition Eq.(\ref{nor}), the physical content of $X_i$ is obvious, it is an overlapping integral over normalized wave functions of initial and final mesons, so we have $0<X_i<1$. When there is no recoil, that is, if $M_f=M$, $f_1 = g_5$ and ${q_{f}}_{_{P_f\bot}} \to {q}_{_{P\bot}}$, we will obtain the largest value $X_i \to 1$, in all other cases, $X_i < 1$. If the two wave functions are much different, then their overlapping will be small, lead to a small $X_i$.

The quantity $X_i$ is almost independent of the initial and final masses, and its physical meaning is obvious, but the definition in Eq.(\ref{X}) is model dependent and non-relativistic, it is not easy to be used by other models. So we will not use it to do calculation, but choose another definition which can be used widely. From the last relation in Eq.(\ref{nonamp}), we can give a equivalent definition
\begin{equation}\label{X1}
X_i=\sqrt{\frac{8\pi\Gamma_i f^2_K M M_f}{{\vec P_f}^3(M+M_f)^2}}.
\end{equation}
This definition is model independent and can be used by all the theoretical models as well as the experiment. From the equations Eq.(\ref{nonamp}), Eq.(\ref{X}) and Eq.(\ref{X1}), we conclude that the quantity $X_i$ is also almost free from the initial and final masses. By using this value, all the theoretical results as well as the experimental data are comparable to each other no matter what initial state mass is used. Another benefit is, $X$ can be used to and may be good at the not well established new state, whose mass and width are not well measured, because $X$ can be used to check the reasonableness between the mass and the corresponding width of the new state.

Eq.(\ref{X}) show that $X$ is almost mass independent, the ratio $\Gamma /{|{\vec P_f}|^3}$ is slightly depend on the mass since we have the relation
\begin{equation}
\frac{\Gamma}{{\vec{P_f}}^3}= \frac{(M+M_f)^2}{8\pi f^2_K M M_f}~X^2.
\end{equation}

\section{Numerical results and discussions}
In our calculation, we solve the full Salpeter equations for the $0^-$ and $1^-$ states to obtain the relativistic wave functions we use to calculate the decay properties.
The interaction kernel in Slapeter equation include a Coulomb vector potential from gluon exchange, a linear confining interaction and a free parameter $V_0$.
In solving Salpeter equation, the following well-fitted parameters \cite{fuhuifeng,Wang:2013lpa} are used:
\begin{center}
 $m_c=1.62\ {\rm GeV}$\;, \ \ \ $m_s=0.5\ {\rm GeV}$\;, \ \ \ $m_d=0.311\ {\rm GeV}$\;, \ \ \ $m_u=0.305\ {\rm GeV}$\;,
\end{center}
then the radial wave functions for $1^-$ vectors $D^*(2007)^0$ and $D^*(2010)^+$ as well as the first radial excited $0^-$ pseudoscalar $D_{s0}(2S)^+$ are obtained \cite{Kim:2003ny,Wang:2005qx}. {In the same time, the corresponding masses of mesons are also obtained, but to our experience, our method tend to give a smaller mass splitting for the $0^-$ state, see Ref.\cite{spectrum} for example, so we will not use this method to predict the mass of $D_{s0}(2S)^+$,  and simply adjust the free parameter $V_0$ in potential to fitting mass data at $2591$ MeV and generate the wave functions.}

\subsection{The decay width of $D^*(2010)^+$}

To confirm our method, we first calculate the strong decays $D^*(2010)^+\to D^0\pi^+$ and $D^*(2010)^+\to D^+\pi^0$, for the later there is an extra parameter 0.5 in the decay width. The results are
\begin{equation}\Gamma(D^*(2010)^+\to D^0\pi^+)=47.5~ {\rm keV},\end{equation}
\begin{equation}\Gamma(D^*(2010)^+\to D^+\pi^0)=20.4~ {\rm keV},
\end{equation}
which are close to the experimental data $\Gamma_{\rm{ex}}(D^*(2010)^+\to D^0\pi^+)=56.5 \pm 1.6$ keV and $\Gamma_{\rm{ex}}(D^*(2010)^+\to D^+\pi^0)=25.6 \pm 1.0$ keV listed in PDG \cite{pdg}.

Now we check the quantity $X$ we have introduced. we define two $Xs$ for the channels $D^*(2010)\to D^0\pi^+$ and $D^*(2010)\to D^+\pi^0$, the results
\begin{equation}\label{18}X({D^0\pi^+})=\sqrt{\frac{24\pi\Gamma({D^0\pi^+}) f^2_{\pi} MM_f }{|{\vec P_f}|^3(M+M_f)^2}}=0.493,\end{equation}
\begin{equation}\label{19}X({D^+\pi^0})=\sqrt{\frac{48\pi\Gamma({D^+\pi^0}) f^2_{\pi} MM_f }{|{\vec P_f}|^3(M+M_f)^2}}=0.484\end{equation}
are very close to experimental data $X_{\rm {ex}}({D^0\pi^+})=0.538\pm0.007$ and $X_{\rm {ex}}({D^+\pi^0})=0.542\pm0.010$ \cite{pdg}.

\subsection{The properties of $D_{s}(2^1S_0)^+$}

After confirm the validity of the method by the decays $D^*(2010)^+\to D\pi$, we apply it to the calculation of $D_{s}(2^1S_0)^+$. To compare with experimental data, we fit the mass of $D_{s}(2^1S_0)^+$ at $2591$ MeV, and the two-body strong decays widths are calculated, the results are
\begin{equation}\Gamma(D_{s0}(2590)^+\to D^*(2007)^0K^+)=10.4 ~\rm{MeV}, \end{equation}
\begin{equation}\Gamma(D_{s0}(2590)^+\to D^*(2010)^+K^0)=9.29 ~\rm{MeV}, \end{equation}
their ratio is $$\frac{\Gamma(D_{s0}(2590)^+\to D^*(2007)^0K^+)}{\Gamma(D_{s0}(2590)^+\to D^*(2010)^+K^0)}=1.12.$$
The full width can be estimated as the sum of them
\begin{equation}\Gamma(D_{s0}(2590)^+) \simeq 19.7 ~\rm{MeV}. \end{equation}

Our results and other theoretical predictions as well as the experimental data are shown in Table III, where we can see, our
prediction is the smallest one, and much smaller than the experimental data $\Gamma_{\rm ex}=89\pm16\pm12$ MeV. In this Table, at first sight, three of the theoretical width predictions at $76\sim78$ MeV are consistent with data, but it is not true, because the used masses of $D_{s}(2^1S_0)^+$ in theoretical models are much larger than data, at least $55$ MeV higher. We have pointed out that the decay width is very sensitive to the mass because the decay happens closing to the threshold. So with different initial masses as input, the decay results are incomparable, that is, it make no sense to directly compare the widths. If alter the initial state mass to the experimental data, these consistent results will become inconsistent, and will be much smaller than data. The reason we get the minimum width is also because the mass we used is the smallest.

\begin{table}
\begin{center}
\caption{Mass, strong decay width of $D_s(2^1S_0)$, recoil momentum ${|{\vec P_f}|}$ (MeV), the ratio ${\Gamma}/{{|\vec P_f|}^3}$ (MeV$^{-2}$) and the model independent quantity $X$.}
\label{tab3}
\begin{tabular}{|c|c|c|c|c|c|c|c|c|c|c|}\hline\hline
&   ours   & \cite{GM}&\cite{liux} & \cite{close}  & \cite{colangelo} &\cite{zhangal}&\cite{wangzh} &Ex~\cite{Aaij:2020voz}\\ \hline						
$M_{D_{s}(2 ^1S_{0})}$  &2591 & 2673&2646 & 2670 &2643 &2650 & 2641 &2591$\pm$6$\pm$7\\\hline
$\Gamma(D_{s}(2S)\to D^* K)$ &19.7 & 76.3& 76.06 & 126 & 33.5 & 78&49,~36 & 89$\pm$16$\pm$12 \\\hline
${|{\vec P_f}|}$ & 270 & 385&350 & 381 &  346 &356 &344  &$270^{+20}_{-22}$ \\\hline
${\Gamma \cdot 10^6}/{|{\vec P_f}|}^3$ &1.01 & 1.34 &1.77& 2.27 & 0.81  &1.74 &1.21,~0.888 &$4.54^{+0.25}_{-0.52}$ \\\hline
~$X=\sqrt{\frac{8\pi\Gamma/2 f^2_K M M_f}{{|{\vec P_f}|}^3(M+M_f)^2}}$~&~0.275~&~0.316~&~0.364~&~0.412~&~0.246~&~0.361~ &~0.301,~0.258~&~$0.585^{+0.015}_{-0.035}$~\\\hline
 \hline
\end{tabular}
\end{center}
\end{table}

In the decay modes of $D_{s}(2^1S_0)^+$, we have the relation $\Gamma \varpropto |\vec{P_f}^3|$, which also indicate that the decay width heavily dependent on the initial state mass, and we pointed out that the ratio $\Gamma/|\vec{P_f}^3|$ can cancel partly the influence of different input masses. So a line of ${\Gamma \cdot 10^6}/{|{\vec P_f}|}^3$ is added in Table III, where in calculation of $|{\vec P_f}|$ and later the quantity of $X$, the averages $M_f\equiv M_{D^*}=(M_{D^*(2007)^0}+M_{D^*(2010)^+})/2$ and $M_{K}=(M_{K^+}+M_{K^0})/2$ are used.  The results confirm our argument, that we can compare the ratios ${\Gamma \cdot 10^6}/{|{\vec P_f}|}^3$ instead of widths no matter what initial masses are used.

When comparing the ratios in Table III, the conclusion is much different from the comparison of decay widths which will result in a wrong conclusion. Our result ${\Gamma \cdot 10^6}/{|{\vec P_f}|}^3=1.01$ MeV$^{-2}$ is not the smallest one, larger than $0.81$ MeV$^{-2}$ in Ref. \cite{colangelo} and $0.888$ MeV$^{-2}$ in Ref. \cite{wangzh}. The results of Refs. \cite{GM,liux,colangelo}, whose widths consist well with data at first sight, are $1.34,~1.77$ and~$1.74$ MeV$^{-2}$, the first one become difference from other two, and all are much smaller than experimental data $4.54^{+0.25}_{-0.52}$ MeV$^{-2}$. Except the value ${\Gamma \cdot 10^6}/{|{\vec P_f}|}^3=2.27$ MeV$^{-2}$ by Ref.  \cite{close} which give the biggest width $\Gamma=126$ MeV, the experimental ratio is much larger than all other theoretical predictions, means they are inconsistent with each other.

The ratio ${\Gamma \cdot 10^6}/{|{\vec P_f}|}^3$ in Table III show us the disagreement between the theoretical calculations and experimental detection, but it can not tell us whether these results as the decay of $D_s(2S)$ state are reasonable or correct, while the model independent quantity $X$ can
realize this purpose, so we calculate the quantity $X$ and add a line in Table \ref{tab3} to show the values of $X$.
Since we usually only have the total decay width from other models and experiment, we will not show the two $Xs$ for the two decay channels, instead we give an average $X$ in Table III and IV using the full width and final average masses as input, where we suppose $\Gamma= \Gamma_{D^{*0}K^+}+\Gamma_{D^{*+}K^0}\simeq 2\Gamma_{D^{*0}K^+}\simeq 2\Gamma_{D^{*+}K^0}$, so here $X\equiv X_{D^{*0}K^+}\equiv X_{D^{*+}K^0}$.  Our result $X=0.275$ consist with $0.316$ in Ref. \cite{GM} and $0.301$ in Ref. \cite{wangzh}, is about half of the experimental data $0.585^{+0.015}_{-0.035}$ \cite{Aaij:2020voz}. We also note that, though there are discrepancies between theoretical predictions, all the theoretical results are smaller than data.

Beside the advantage that it is model independent, we point out that quantity $X$ has another more convenient advantage, that it can be used to compare the results between similar but different decay channels, for example, we can compare the results of decays $D_s(2^1S_0) \to D^*K$ and $D^*(2010)\to D\pi$. The conclusion is the quantity $X$ of the former will be much smaller than those of the later, because, (1) the radial wave functions for $D_s(2^1S_0)$ and $D^*(1^3S_1)$ in the decay $D_s(2^1S_0) \to D^*K$ are much different, one is $2S$ state, another is $1S$ state; while in the decay $D^*(2010)\to D\pi$, both $D^*(2010)$ and $D(1^1S_0)$ are $1S$ state, their radial wave functions are  equal in the non-relativistic limit; so the overlapping between $D_s(2^1S_0)$ and $D^*(1^3S_1)$ will be much smaller than those between $D^*(2010)$ and $D(1^1S_0)$; (2) more important, there is a nodal structure in the $2S$ wave function, contributions from the two sides of the node are cancelled, which will result in a small $X$ for the decay $D_s(2^1S_0) \to D^*K$; (3) we will show later that large $|\vec P_f|$ will depresses the $X$ value, the $|\vec P_f|\simeq 300$ MeV in decay $D_s(2^1S_0) \to D^*K$ is much larger than $|\vec P_f|= 39$ MeV in $D^*(2010)\to D\pi$. So with these three comments, compared with $X_{D^*(2010) \to D\pi}$, we should obtain a much smaller $X_{D_s(2S) \to D^*K}$, but currently the experimental data are $X_{D^*(2010) \to D\pi}=0.540\pm0.009$ and $X_{D_s(2S) \to D^*K}=0.585^{+0.015}_{-0.035}$. Since $D^*(2010)$ is well established, we conclude that $X_{D_s(2S) \to D^*K}=0.585^{+0.015}_{-0.035}$ is too big to be a reasonable value for the transition $D_s(2^1S_0) \to D^*K$, it should be much smaller like our result which is about half of the current data.

The unreasonable conflicting data $X_{D_s(2S) \to D^*K}=0.585^{+0.015}_{-0.035}$ indicates that the current detected mass and full width of $D_{s0}(2590)^{+}$ supposed as state $D_s(2^1S_0)^+$ do not match well to each other. To obtain a rational $X_{D_s(2S) \to D^*K}$ which should be much smaller than current data, the full width $89\pm16\pm12$ MeV is too broad with the low mass $2591\pm6\pm7$ MeV, or the mass $2591\pm6\pm7$ MeV is too low with current broad width. So according to value of $X$, we can not conclude that the $D_{s0}(2590)^{+}$ is the pure state $D_s(2^1S_0)^+$.

\subsection{The Character of $X$}

\begin{table}
\begin{center}
\caption{Dependence of the decay width $\Gamma$ (MeV), ratio ${\Gamma \cdot 10^6}/{|{\vec P_f}|^3}$ (MeV$^{-2}$) and quantity $X$ on the variation of the $D_s(2S)$ mass (MeV) or the recoil momentum ${|{\vec P_f}|}$ (MeV).}
\label{tab4}
\begin{tabular}{|c|c|c|c|c|c|c|c|c|}\hline\hline						
~~$M_{D_{s}(2 ^1S_{0})}$~~&~~2600~~&~~2610~~&~~2620~~&~~2630~~&~~2640~~&~~2650~~&~~2660~~&~~2670~~\\\hline
~~${|{\vec P_f}|}$~~&~~284~~&~~299~~&~~314~~&~~328~~&~~342~~&~~356~~&~~369~~&~~381~~ \\\hline
~~$\Gamma(D_{s}(2 S)\to D^* K)$~~&~~22.9~~&~~26.4~~&~~30.0~~&~~33.8~~&~~37.6~~&~~41.5~~&~~45.5~~&~~49.5~~ \\\hline
~~${\Gamma \cdot 10^6}/{|{\vec P_f}|^3}$~~&~~1.00~~ &~~0.984~~&~~0.968~~&~~0.955~~&~~0.939~~&~~ 0.923~~&~~0.909~~&~~0.893~~\\\hline
~~$X=\sqrt{\frac{4\pi\Gamma f^2_K M M_f}{|{\vec P_f}|^3(M+M_f)^2}}~~$~~&~~0.274~~&~~0.272~~&~~0.269~~ &~~0.268~~&~~0.265~~&~~0.263~~&~~0.261~~&~~0.259~~\\\hline
 \hline
\end{tabular}
\end{center}
\end{table}

In Table \ref{tab4}, we vary the input initial state $D_s(2^1S_0)^+$ mass from $2600$ to $2670$ MeV, and show the corresponding variations of other physical quantities. $|{\vec P_f}|$ changes from $284$ to $381$ MeV, it is very sensitive, but the most sensitive quantity is the decay width $\Gamma$, increases from $22.9$ to $49.5$ MeV. While the ratio ${\Gamma \cdot 10^6}/{|{\vec P_f}|^3}$ and quantity $X$ decrease slightly along with the increasing mass. ${\Gamma \cdot 10^6}/{|{\vec P_f}|^3}$ decreases from $1.0$ to $0.893$ MeV$^{-2}$, $X$ from $0.274$ to $0.259$, as expected they are very stable along with the variation of mass, which indicate that their dependence on mass is removed to a great extent, especially the quantity $X$. So as we pointed out, this character of independence on mass make the $X$ suitable in dealing with a not well established new state, since usually its mass has large uncertainties which may result in large errors in the calculation of decays or productions, while $X$ is almost mass independent, then despite the large errors of mass, we can obtain a useful result. Further, because the $X$ indicate the overlap of integral over initial and final state wave functions, so physically it has a reasonable range, which can tell us the theoretical calculation is correct or not.

Table \ref{tab4} shows the $D_s(2^1S_0)^+$ mass range from 2590 MeV to 2670 MeV, and the corresponding information of decays to $D^*K$. In fact, if $D_s(2^1S_0)^+$ mass is above 2660 MeV, it can decay to $D^*_s\eta$, and if its mass is as high as 2763 MeV, the $DK^*$ decay channel will be active, but considering that the $D_s(2^3S_1)$ mass is 2710 MeV, it is unlikely that the mass of $D_s(2^1S_0)^+$ is higher than 2710 MeV. Therefore, in the mass range of $2660\sim2710$ MeV, due to the small $|{\vec P_f}|$ in this case, the contribution of $D^*_s\eta$ will be very small, for example, Close {\it et al.} \cite{close} show us that $\Gamma(D_{s}(2 S)\to D_s^* \eta)=0.5$ MeV when $m_{D_s(2^1S_0)}=2670$ MeV. So we did not show $D^*_s\eta$ channel in Table \ref{tab4}.

\subsection{Decaying $D(2^1S_0)^0\to D^*\pi$}

We also calculate the strong decays of $D(2^1S_0)^0$, whose candidate is the new particle $D_0(2550)^0$. Choosing the average mass 2564 MeV in PDG \cite{pdg} as input, we obtain \begin{equation}\Gamma(D_0(2550)^0\to D^{*+}\pi^-)=21.2~\rm MeV,\end{equation}
\begin{equation}\Gamma(D_0(2550)^0\to D^{*0}\pi^0)=11.0~\rm MeV.\end{equation}
With mass $2564\pm20$ MeV, $D_0(2550)^0$ also has $D(2550)\to D_0^{*}(2400)\pi$ and $D(2550)\to D_1^{*}(2420)\pi$ decay channels where a $P$ wave is involved in, but unlike the decays $D(2550\to D^{*}\pi$, these two channels have very small phase spaces, then their contributions to the whole decay width are tiny \cite{GM,wangzh} and can be ignored. In previous paper Ref.\cite{wangzh}, two different amplitude formulas are used, and we got little larger total decay width, 43 MeV or 46 MeV, but all these theoretical results of $D(2^1S_0)^0$ decay widths are much smaller than the average value $\Gamma_{\rm{ex}}(D(2550))=135\pm17$ MeV in PDG \cite{pdg}.

The corresponding quantity $X$ is
\begin{equation}X_{D(2^1S_0) \to D^*\pi}=\sqrt{\frac{16\pi\Gamma f^2_{\pi} M M_f}{3|{\vec P_f}|^3(M+M_f)^2}}=0.143,\end{equation}
in previous paper Ref.\cite{wangzh}, it is 0.167 or 0.177, all are much smaller than current data $X^{ex}=0.292^{+0.035}_{-0.033}$. We find that both experimental data and theoretical results show us that $X_{D(2^1S_0) \to D^*\pi}$ is much smaller than $X_{D^*\to D\pi}$, this relation is correct and reasonable, because the former has the nodal structure in wave function, see Sec.V.B. Theoretically, $X^{th}_{D(2^1S_0) \to D^*\pi}$ is much smaller than $X^{th}_{D_s(2^1S_0) \to D^*K}$, this may be mainly due to the difference of the phase spaces, the recoil momentum $|{\vec P_f}|=480$ MeV for the former is much larger than $|{\vec P_f}|=270$ MeV for the later. Although the decay width is enhanced, the quantity $X$ is depressed by the large recoil momentum, see Eq. (\ref{X}) and the results in Table \ref{tab4}.

\subsection{Conclusions}

We choose the Reduction Formula, PCAC and Low Energy Theory to reduce the $S$ matrix of a two-body OZI allowed strong decay, avoid using the wave function of light $K$ meson, the covariant transition amplitude is written as overlapping integral over the relativistic wave functions of the initial and final heavy mesons, where the relativistic wave functions are obtained by solving the full Salpeter equations.

We first study the strong decays of $D^*(2010)$, the predicted
$\Gamma(D^0\pi^+)=47.5$ keV and $\Gamma(D^+\pi^0)=20.4$ keV are close to the experimental data $56.5 \pm 1.6$ keV and $25.6 \pm 1.0$ keV \cite{pdg}.
We studied the overlap integral over wave functions of initial and final states and it is parameterized as quantity $X$, our theoretical results
$X_{D^0\pi^+}=0.493$ and $X_{D^+\pi^0}=0.484$
consist with experimental data $0.538\pm0.007$ and $0.542\pm0.010$ \cite{pdg}.
These studies confirm the validity of this method and the quantity $X$.

We then study the newly observed $D_{s0}(2590)^{+}$ as the $D_s(2^1S_0)^+$ candidate.
We find the detected mass $2591$ MeV is smaller than all the theoretical predictions, at least several tens of MeV. According to the mass splittings detected in experiments, the expected mass of $D_s(2^1S_0)^+$ is located at $2620 \to 2665$ MeV. Using the mass $2591$ MeV, the calculated width $\Gamma(D_{s0}(2590)^+) \simeq 19.7$ MeV is much smaller than data
$\Gamma_{\rm{ex}}=89\pm16\pm12$ MeV.

We find that the decay width $\Gamma(D_{s}(2 S))$ is highly sensitive to its mass, while the ratio $\Gamma/{|{\vec P_f}|^3}$ and quantity $X$, especially the later, are almost mass independent, so for a state with not well measured mass, instead of width, these two stable quantities are much useful. All the theoretical predictions of $\Gamma/{|{\vec P_f}|^3}$  are smaller than experimental data. To check the reasonableness of the results, the $X$ value is needed, comparing the quantities $X_{D_s(2S) \to D^*K}=0.25\sim 0.41$ in theory, and $X^{ex}_{D^*(2010) \to D\pi}=0.540\pm0.009$ in experiment, the current data $X^{ex}_{D_s(2S) \to D^*K}=0.585^{+0.015}_{-0.035}$ is too big to be a reasonable value.

We conclude that the current mass and width of $D_{s0}(2590)^{+}$ in experiment as the candidate of $D_s(2^1S_0)^+$ do not match to each other, so it is too early to say $D_{s0}(2590)^{+}$ is the state $D_s(2^1S_0)^+$. We have pointed out that the detection channel, $D_{s0}(2590)^+\to D^+K^+\pi^-$, is a direct three-body decay process, not a cascade decay of a two-body decay channel. Therefore, it is very important in experiment to detect the two-body processes $D_{s0}(2590)^+\to D^*(2007)^0K^+$ and $D_{s0}(2590)^+\to D^*(2010)^+K^0$, study their decay widths, branching ratios, which is necessary to reveal the nature of $D_{s0}(2590)^+$, and shed light on the confusion whether it is the state $D_s(2^1S_0)^+$.

\vspace{0.7cm} {\bf Acknowledgments}

This work was supported in part by the National Natural Science Foundation of China(NSFC) under the Grants Nos. 12075073, 12075074, 11865001, and the Natural Science Foundation of Hebei province under the Grant No. A2021201009.

\begin{appendix}
\section{The covariant transition amplitude}

According to the Mandelstam formalism \cite{Mandelstam}, the transition  amplitude can be written as an overlapping integral over the initial and final states' Bethe-Salpeter relativistic wave functions,
\begin{equation}\label{ha-amplitude}
\begin{aligned}
\langle {D^*}(P_f)^0|\bar{q}\gamma_{\mu}\gamma_5s|D_{s0}(P)^+\rangle
=\int \frac{\mathrm{d}^{4} q}{(2 \pi)^{4}} \operatorname{Tr}\left[\overline{\chi}_{_{P_{f}}}\left(q_{f}\right)
S^{-1}\left(p_{c}\right) \chi_{_P}(q) \gamma_{\mu}\gamma_5\right],
\end{aligned}
\end{equation}
where $\chi_{_P}(q)$ and $\chi_{_{P_{f}}}(q_{f})$ are the relativistic BS wave functions of initial and final mesons, $S(p_c)$ is the propagator. The relation between the initial relative momentum and final relative momentum is
$q_f=q+\frac{m_c}{m_c+m_s} P-\alpha'_c P_f.$

The wave function $\chi_{_P}(q)$ is the solution of BS equation,
\begin{equation}
\label{BS}
\chi_{_P}(q)
=iS(p_c)\int{\frac{d^4k}{(2\pi)^4}V(P,k,q)
\chi_{_P}(k)}S(-p_s),
\end{equation}
where $V$ is the interaction kernel between quark and antiquark.
In the condition of instantaneous approximation, the kernel $V$ will only depend on the three dimensional quantity $q_{_{P\bot}}-k_{_{P\bot}}$. With the following definitions,
$$\varphi(q_{_{P\perp}}) \equiv i \int{\frac{dq_{_P}}{2\pi} \chi_{_P}(q)},
\qquad
\eta_{_P}(q_{_{P\perp}}) \equiv \int{\frac{dk^3_{_{P\perp}}}{(2\pi)^3} V(k_{_{P\perp}},q_{_{P\perp}})\varphi(k_{_{P\perp}}) }.$$
the BS equation Eq.(\ref{BS}) can be written as $
\chi_{_P}(q)=S(p_c)\eta_{_P}(q_{_{P\perp}})S(-p_s).
$
Then Eq.(\ref{ha-amplitude}) becomes to
\begin{equation}
\begin{aligned}\int \frac{\mathrm{d}^{4} q}{(2 \pi)^{4}} \operatorname{Tr}\left[S(-p_u)\overline{\eta}_{_{P_{f}}}({q_{f}}_{_{P_f\bot}})
 S(p_c)\eta_{_P}(q_{_{P\bot}})S(-p_s) \gamma_{\mu}\gamma_5\right],
\end{aligned}
\end{equation}
where the propagator is function of projection operator
\begin{equation}
\begin{aligned}
&{(-1)^{J+1}}iS({(-1)^{J+1}}p_{_J}) = \frac{\Lambda^+((-1)^{J+1}p_{_{JP\perp}})}{(-1)^{J+1}p_{_{JP}}-\omega_{_J}+i\epsilon} + \frac{\Lambda^-((-1)^{J+1}p_{_{JP\perp}})}{(-1)^{J+1}p_{_{JP}}+\omega_{_J}-i\epsilon},
\end{aligned}
\end{equation}
with
$\Lambda^\pm((-1)^{J+1}p_{_{JP\perp}})=\frac{1}{2\omega_{_J}}\Big[\frac{\slashed P}{M}\omega_{_J} \pm((-1)^{J+1}m_{_J}+\slashed p_{_{JP\perp}})\Big],$ $
\omega_{_J}\equiv\sqrt{m_{_J}^2-p^2_{_{JP\perp}}}
$, $J=1$ for a quark, and $J=2$ for an anti-quark.

If we omit the terms with negative projection operators $\Lambda^-$s, whose contributions are very small and are neglected \cite{Kim:2003ny}, then the transition amplitude can be written as
$$\int \frac{\mathrm{d}^{3} q_{_{P\bot}}\mathrm{d} q_{_{P}}}{(2 \pi)^{4}} \operatorname{Tr}\left[\frac{\Lambda^+(-p_{u_{P\perp}})}
{p_{u_P}+\omega_{u}-i\epsilon}
{\frac{\not\!P_f}{M_f}}\widetilde\Lambda^{+}({-p_{u}}_{P_{f\bot}})
\overline{\eta}_{_{P_{f}}}({q_{f}}_{P_f\bot})\widetilde\Lambda^{+}({p_{c}}_{P_{f\bot}})
\right.$$
\begin{equation}
\begin{aligned}
\times\left.
{\frac{\not\!P_f}{M_f}}
\frac{\Lambda^+(p_{c_{P\perp}})}{p_{c_P}-\omega_{c}+i\epsilon}
\eta_{_P}(q_{P\bot})\frac{\Lambda^+(-p_{s_{P\perp}})}
{p_{s_P}+\omega_{s}-i\epsilon} \gamma_{\mu}\gamma_5\right],
\end{aligned}
\end{equation}
where $\left({\frac{\not\!P_f}{M_f}}\right)^2=1$ is inserted twice, and the relations $\frac{\not\!P_f}{M_f}=\widetilde\Lambda^{+}({\pm p_{i}}_{P_{f\bot}})+
\widetilde\Lambda^{-}({\pm p_{i}}_{P_{f\bot}})$ are used with $i=c,u$; $\tilde{x}$ means that the quantity $x$ is written in the center of mass system of the final state.

Finally, after finishing the contour integral over $q_{_{P}}$, we obtain the covariant expression for the transition amplitude shown in Eq. (\ref{amplitude1}), in which the following Salpeter equation \cite{Sal,Kim:2003ny}
\begin{equation}
\begin{aligned}
{\varphi}^{++}_{_{P}}({q}_{_{P\bot}})=\frac{\Lambda^+(p_{c_{_{P\perp}}})
\eta_{_P}(q_{_{P\bot}}){\Lambda^+(-p_{s_{P\perp}})}}{M-\omega_c-\omega_{s}}
 \end{aligned}
\end{equation}
is used.

\end{appendix}

\end{document}